\documentclass[aps,prstper,preprint]{revtex4-1} 

\usepackage{graphicx}
\usepackage{dcolumn}
\usepackage{bm}
\usepackage{color}
\usepackage{amsmath}
\usepackage{framed}

\definecolor{red}{rgb}{1,0,0}
\definecolor{orange}{rgb}{1,0.5,0}
\definecolor{green}{rgb}{0.13,0.55,0.13}
\definecolor{purple}{rgb}{0.5,0,1}

\begin{document}

\preprint{Caballero, et al.}

\title{Implementing and assessing computational modeling in introductory mechanics}

\author{Marcos D. \surname{Caballero}}
	\altaffiliation[Current Address: ]{Department of Physics, University of Colorado at Boulder, Boulder, CO 80309}
\author{Michael F. \surname{Schatz}}
	\email[Corresponding Author: ]{michael.schatz@physics.gatech.edu}
\affiliation{Center for Nonlinear Science and School of Physics, Georgia Institute of Technology, Atlanta, GA 30332}
\author{Matthew A. \surname{Kohlmyer}}
	\altaffiliation[Current Address: ]{Advanced Instructional Systems, Inc., Raleigh, NC 27696}
\affiliation{Department of Physics, North Carolina State University, Raleigh, NC 27695}

\date{\today}

\begin{abstract}
Students taking introductory physics are rarely exposed to computational modeling. In a one-semester  large lecture introductory calculus-based mechanics course at Georgia Tech, students learned to solve physics problems using the VPython programming environment. During the term, 1357 students in this course solved a suite of fourteen computational modeling homework questions delivered using an online commercial course management system. Their proficiency with computational modeling was evaluated with a proctored assignment involving a novel central force problem. The majority of students (60.4\%) successfully completed the evaluation. Analysis of erroneous student-submitted programs indicated that a small set of student errors explained why most programs failed. We discuss the design and implementation of the computational modeling homework and evaluation, the results from the evaluation, and the implications for computational instruction in introductory STEM courses.
\end{abstract}

\maketitle

\section{\label{sec:vpintro}Introduction}

Computation (the use of a computer to numerically solve, simulate, or visualize a physical problem) has revolutionized scientific research and engineering practice. 
In science and engineering, computation is considered to be as important as theory and experiment.\cite{siamstatementweb} 
Systems that cannot be solved in in closed form are probed using computation; experiments that are impossible to perform in a lab are studied numerically.\cite{fenton2005modeling,gonzalez2009black} 
Yet, as computer usage has grown (e.g., facilitation of homework), most introductory courses have failed to introduce students to computation's problem solving powers.

Computer usage (as opposed to {\it computation}) is ubiquitous in large introductory courses, and much work has been done to understood about the role of computers in these courses. \cite{morote2009course} 
By contrast, little is known about the role that computers might play as problem-solving tools in introductory courses.
For example, given the constraints of a traditional large lecture STEM course, is it possible to provide sufficient instruction in computational problem solving techniques such that students can apply these tools to new problems?
Moreover, what challenges do students face when learning and applying computational problem solving techniques, and how might we mitigate those challenges through instruction?  In this paper, we describe our efforts to address these questions.

Using computation in introductory physics courses has several potential benefits. 
Students can engage in the modeling process to make complex problems tractable. 
Students can also use computation to explore the applicability and utility of physical principles. 
In a way, students who compute are participating in work that is more representative of what they will do as professional scientists and engineers.\cite{macdonald1988muppet,redish1993student,schecker1993learning,schecker1994system} 
When constructing simulations, students are constrained by the programming language to certain syntactic structures. 
Hence, they must learn to contextualize problems in a way that produces a precise representation of the physical model.\cite{disessa1986boxer,sherin1993dynaturtle} 
One of computation's key strengths lies in its utility for visualizing and animating solutions to problems. 
These visualizations can improve students' conceptual understanding of physics.\cite{finkelstein2006high}
It is important to distinguish between computational modeling and simply writing program statements. Computational modeling is the expressing, describing, and/or solving of a science problem by using a computer. It requires additional steps beyond simple programming, such as contextualizing the programming task and making sense of the physical model. While programming is taught in most introductory computer science courses, computational modeling is not.

We used computation in a large-enrollment introductory calculus-based mechanics course at the Georgia Institute of Technology to develop students' modeling and numerical analysis skills. 
We built upon previous attempts to introduce computation in introductory physics laboratories \cite{computajp,beichner2010labs} by extending its usage to other aspects of students' coursework.
In particular, we taught students to use the VPython programming environment to construct models that predict the motion of physical systems. \cite{vpythonWebsite}
We describe the design and implementation of homework problems to develop students' computational modeling skills in a high-enrollment foundational physics course (Sec. \ref{sec:vpdeploy}). 
We also provide the first evaluation and explication of students' skills when they attempt individually to solve a novel computational problem in a proctored environment (Secs. \ref{sec:eval}--\ref{sec:cluster}). 
We discuss implications for instructional design, considerations regarding student epistemology and the assessment of knowledge transfer as well as the broader implications of teaching computation to introductory physics students (Sec. \ref{sec:closing}).

\section{Approaches to implementing computation}\label{sec:comp-approaches}

Since the development of inexpensive modern microcomputers with visual displays, there have been a number of attempts to introduce computation into physics courses. We review these attempts by decomposing them along two dimensions ({\it openness of the environment} and {\it size of intended population} ) to indicate how our approach fits in with previous work. 

Some have worked closely with a small number of students to develop computational models in an {\it open computational environment}. 
Historical examples include the Maryland University Project in Physics and Educational Technology, \cite{macdonald1988muppet,redish1993student} STELLA, \cite{schecker1993learning,
schecker1994system} and the Berkeley BOXER project. \cite{disessa1986boxer,sherin1993dynaturtle}
Open computational environments are analogous to ``user-developed'' code in scientific research (as opposed to prepackaged or ``canned''software). 
Students who learn to use an open environment have the advantage of viewing and altering the underlying algorithm on which the computational model depends. 
Moreover, students using an open computational environment might learn to develop new models to solve new problems.
It is true, however, that students must devote time and cognitive effort to learning the syntax and procedures of the open environment's programming language.  It is most desirable to have students focus
on developing the physical model without spending excessive
time and effort on the details of constructing working code (e.g., message handling, drawing graphics, garbage collection). 
It is, therefore, important to consider students' experience (or lack thereof) with programming when choosing an open computational environment. 

Others have developed {\it closed computational environments} for use at a variety of instructional levels. 
These environments have been deployed in a number of settings ranging from a few students to large lecture sections.
Examples of closed environments include Physlets \cite{christian1998developing} and the University of Colorado's Physics Educational Technology simulations. \cite{perkins2006phet, wieman2006powerful}
Closed computational environments are analogous to ``canned'' code in scientific research. 
Students can set up and operate the program but do not construct it, nor do they have access to the underlying model or modeling algorithm (a ``black box'' environment). 
User interaction in closed computational environments is often limited to setting or adjusting parameters; a small number of Physlets do allow users to manipulate equations of motion. 
Closed computational environments are useful because they typically require no programming knowledge to operate, run similarly on a variety of platforms with little more than an Internet browser, and produce highly visual simulations. 

It is possible for computational models created in any open environment to be used in a closed manner. 
Users can be restricted (formally or informally) from viewing or altering the underlying model. 
Models developed using Easy Java Simulations \cite{esquembre2004easy} (EJS), an open-source, freely available Java simulation authoring tool, have been used in a closed manner at a variety scales and instructional levels. \cite{christian2007modeling,belloni2007osp} 
However, since it is an open environment, all the features of the physical and computational models in an EJS simulation are accessible to the user. 
Furthermore, EJS has made authoring high-quality simulations possible for students without much programming experience. Some have proposed teaching upper-divison science majors to develop computational models using EJS. \cite{esquembre2007integrate}

VPython,\cite{vpythonWebsite} an open computational environment, has been used to teach introductory physics students to create computational models of physical phenomena. \cite{computajp}
Typically, students write all the program statements necessary to model the physical system (e.g., creating objects, assigning variables, and performing numerical calculations) using IDLE, VPython's default development environment. 
The additional details of model construction (e.g., drawing graphics, creating windows, mouse interactions) are handled by VPython and are invisible to the students. 
VPython supports full three-dimensional graphics and animation without the additional burden to students of learning a great deal of object-oriented programming. \cite{scherer2000vpython} Given its roots in the Python programming language, VPython can be a powerful foundation for students to start learning the tools of their science or engineering trade.
Moreover, VPython is an open-source, freely available environment accessible to users of all major computing platforms.

The Matter \& Interactions (M\&I) curriculum \cite{mandi1} introduces computational modeling as an integral part of the introductory physics course. 
Many of the accompanying laboratory activities are written with VPython in mind, and a number of lecture demonstrations are VPython programs.
In the traditional implementation of M\&I, the practice of constructing computational models is limited to the laboratory. 
In a typical lab, students work in small groups to complete a computational activity by following a guided handout. 
They pause periodically to check their work with other groups or their teaching assistant (TA).
Students' computational modeling skills are evaluated by solving fill-in-the-blank test questions in which they must write the VPython program statements missing from a computational model.

Our approach to teaching computation uses an open environment (in VPython) and builds on our experience with M\&I to extend the computational experience beyond the laboratory.
We chose to use an open environment to teach computation in order to provide students with the opportunity to look inside the computational ``black-box'' and construct the model themselves.
Furthermore, we aimed to teach students how to develop solutions to analytically intractable problems.
We chose VPython (instead of e.g. Java, C, or Matlab) because it has a number of helpful features for novice programmers, can be used to construct high-quality three-dimensional simulations easily, and is freely available to our students.
VPython is also conveniently coupled to M\&I, allowing us to leverage our years of experience with teaching M\&I.
While our implementation builds on our M\&I experience, it is not limited to that experience.
We describe our implementation philosophy in the next section.

\section{Design and Implementation of Computational Homework}\label{sec:vpdeploy}

We aimed to develop an instructional strategy that helps computation permeate the course, but which does not require that students have previous programming experience. 
At Georgia Tech, half of our introductory mechanics students have no programming experience; the other half have had some introduction to programming in high school or in their previous two semesters at Georgia Tech. \footnote{This data is self-reported. Students were polled in class using a ``clicker'' question.}
The Georgia Tech ``introduction to programming'' course is typically taught using MATLAB, not VPython.
Furthermore, our implementation had to be easily deployable across large lecture sections, the setting in which most introductory calculus-based courses are taught.
Our philosophy held that students should learn computation by altering their own lab-developed programs to solve slightly modified problems outside of lab. 
This design philosophy was informed by what research scientists do quite often; they write a program to solve a problem and then alter that program to solve a different problem that is of interest to them.
We envisioned developing computational activities that would start with guided inquiry and exploration in the laboratory, and finish with independent practice on homework. 
In this vision, students would work with TAs in the laboratory to develop a program that solves a problem. 
Students would then modify that program on their own to solve a different problem on their homework.
In our implementation, eleven out of a total of thirteen mechanics laboratories contained a computational modeling activity. Students solved thirteen computational modeling homework problems (one per week) which were embedded in their regular homework ($\sim$15-21 problems per week).

The class of problems that becomes available to students who have learned computation is large and diverse; we chose to focus our efforts on teaching students to apply Newton's second law iteratively to predict motion.
Students taking a typical introductory mechanics course would learn several motion prediction equations emphasizing kinematics, a way of describing the motion without explicitly connecting changes in the motion to forces (i.e., dynamics).
These kinematic formulas are quite limited in their applicability; students can only apply them to problems in which the forces are constant.
This can confuse students when they are presented with a situation where such formulas do not apply. \cite{kohlmyer_thesis}
Students taking most introductory mechanics courses are limited to solving problems that are analytically tractable. Furthermore, the special case of constant force motion is usually the capstone of motion prediction in an introductory mechanics course. Some courses might teach students to analytically determine the motion of an object subjected to some integrable $a(t)$ or subjected to linear ($F \sim v$) air drag in one dimension, but neither case demonstrates the full predictive power of Newton's second law. 
By contrast, computation allows instructors to start first and foremost with Newton's second law and emphasize its full predictive power. 
Students can numerically model the motion of a system as long as they are able to develop a physical model of the interactions and express it in the computational environment. 
The numerical integration technique used to predict motion is a simple algorithm.

\begin{figure}
\includegraphics[width=\linewidth]{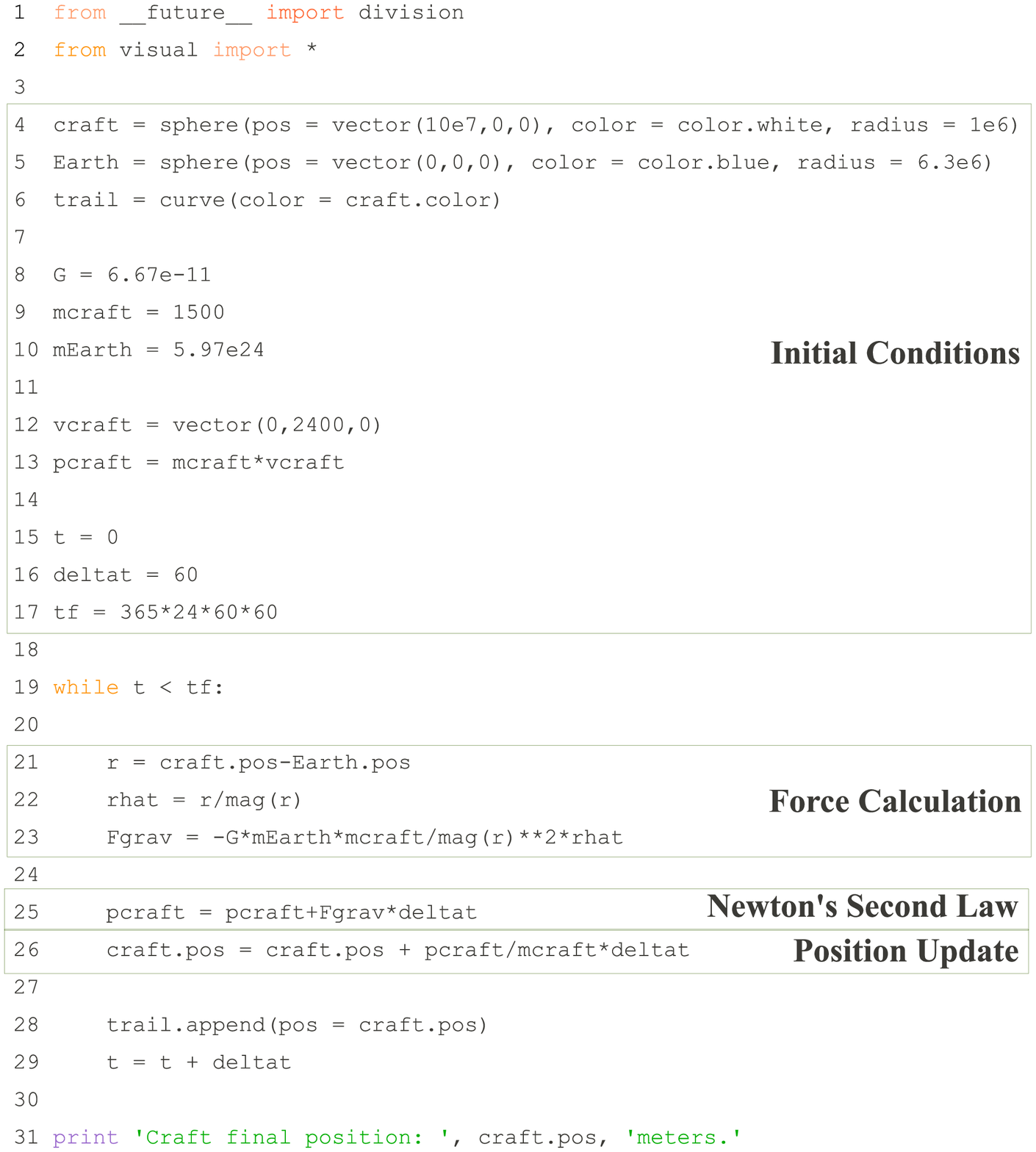}
\caption{[Color] - Under the guidance of their TAs, students wrote the VPython program above in the laboratory. This program modeled the motion of a craft (size exaggerated for visualization) orbiting the Earth over the course of one ``virtual'' year. To construct this model, students had to create the objects and assign their positions and sizes (lines 4--6), identify and assign the other given values and relevant initial conditions (lines 8-10, 12--13 and 15--17), calculate the net force acting on the object of interest appropriately (lines 21--23), and update the momentum and position of this object in each timestep (lines 25--26).}\label{fig:hwsamplecode}
\end{figure}

As a concrete example of our design, we show a mid-semester laboratory activity and homework problem in which students modeled the gravitational interaction between two bodies. 
In this example, students develop a VPython program that models the motion of a craft as it orbits the Earth (Fig. \ref{fig:hwsamplecode}).
Students later make a number of modifications to this program to solve a new problem on their homework.
This example is useful because it illustrates both the level of sophistication we expect of students and the types of alterations we ask students to make on their homework.

In groups of three, students wrote a program in the laboratory to model the motion of low-mass craft as it orbits the Earth (Fig \ref{fig:hwsamplecode}). 
In VPython, they created the objects (lines 4--6), assigned the constants and initial conditions (lines 8-10, 12--13 and 15--17), and set up the numerical integration loop (lines 19--29). 
The program statements in this loop included those which calculated the net force (lines 21--23) and updated the momentum (using Newton's second law) and position of the craft (lines 25--26). 
When developing their physical model, students discussed the motion of the Earth; it experiences the same (magnitude) force as the space craft. 
Students reasoned that its change in velocity is negligible.
Hence, students did not include the motion of the Earth in their VPython program.
When writing this program in the lab, students could seek help from TAs at any time.
The accuracy of the students' completed code was checked by their TAs. 
After completing the lab, students had written a VPython program which could model the motion of the craft moving around the Earth for any arbitrary amount of time.  

In the week following the lab, students solved a computational homework problem in which they used the computational model that they had written in lab to solve a modified problem.
Students were asked to alter their initial conditions to predict the position and velocity of the craft after a specified time.
To solve this problem successfully, students had to identify and make changes to their initial conditions (lines 4, 6, 9--10, and 12) and their integration time (line 17).
In addition, students had to add an additional print statement (after line 31) to print the final velocity of the craft.

Computational homework problems were deployed using the WebAssign course management system, which facilitated the weekly grading of students' solutions. 
To create the homework problem, we numerically integrated several hundred initial conditions and stored the solutions, including final quantitative and qualitative results. 
Each student was assigned a random set of initial conditions corresponding to a particular set of results. 
Randomization ensured a high probability that each student received a unique problem. 
Students used their assigned initial conditions and wrote additional statements to answer the questions posed in the problem. 
Students entered numerical answers into answer fields, and selected check-boxes to answer to qualitative questions.
On these weekly assignments, only students' final results were graded; their code was uploaded for verification purposes, but not graded.
Grading programs for structure and syntax at this large scale requires additional work by TAs who are already charged with a number of other teaching and grading tasks.
Computational homework problems were generally completed in the week that followed the associated laboratory activity.

To facilitate student success and to help students learn to debug their programs, each assignment contained a {\it test case} -- an initial problem for which the solution (i.e., the results from the numerical integration) was given.
When writing or altering any computer program, even experts can create programming errors (bugs).
Learning to debug programs is part of learning how to develop computational models.
This test case ensured that a student's program worked properly and helped to instill confidence in students who might otherwise have been uncomfortable writing VPython programs without the help of their group members or TAs.
After a student checked her program against the {\it test case}, she completed the {\it grading case} -- a problem without a given solution. 

In keeping with our overall design philosophy, most of the homework problems which students solved had similar designs as the aforementioned example. 
In particular, students built a computational model in the laboratory and independently used that fully-functioning model to solve a modified problem on their homework. 
On the first four homework assignments  (of which the previous example is the fourth) students made only a few modifications to their programs, altering their initial conditions and adding a new print statement.
In the next several labs, students learned to model more complicated systems (e.g., three body gravitational problem, spring dynamics with drag) while learning new algorithms such as decomposing the net force vectors into radial and tangential components.
Students also learned to represent these force components as arrows in VPython.
On the homework problems associated with these labs, students still used their lab-developed programs to solve new problems by changing initial conditions and representing new quantities with arrows.

The last two homework problems which students solved were not related to the laboratory exercises; we intended to emphasize the utility of learning to predict motion using Newton's second law. 
To solve these problems, students wrote all the statements missing from a partially-completed code to predict the motion of two interacting objects. 
These were interactions which students had not seen before (e.g., the anharmonic potential and the Lennard-Jones interaction).
In these problems, we omitted the appropriate initial conditions and the statements that numerically integrated the equations of motion.
Students had to contextualize the word problem into a programming task and produce a precise representation of the problem in the VPython programming environment.
With regard to programming tasks, students had to do no more than identify and assign variables and implement the usual motion prediction algorithm for these two problems.
A similarly-designed problem was used as an evaluative assignment and is discussed in detail in Sec. \ref{sec:eval}.

\section{Evaluating computational modeling skills}\label{sec:eval}

Students performed as well on computational modeling homework problems as they did on their analytic homework.
We found no statistical difference between students' performance on their computational homework and their performance on analytic homework using a rank-sum test (Analytic 84.6\% vs. Computational 85.8\%).\cite{conover_nonpara} 
A side effect of the additional practice was an overall increase in student performance on hand-written fill-in-the-blank programming questions on exams.
However, this result did not indicate what fraction of students were able to solve these computational homework problems without assistance. 
While randomizing initial conditions between each student's realization ensured that students' solutions differed with high probability, working programs could be distributed easily from student to student by email.
We note that the distribution of students' programs might not be deleterious; students who received these programs must still have read and interpreted the program statements to enter in their initial conditions, make changes to the force law, or print additional quantities.
This is a more complex interaction than simply plugging numbers into a algebraic solution that they discovered online.
In a sense, students who worked with shared code were using a ``closed'' computational environment.

Nevertheless, we wanted to measure how effective students were at individually solving computational problems.
We delivered a proctored laboratory assignment during the last lab of three different semesters to evaluate students' computational skills on an individual basis. 
Students received a partially-completed program that created two objects (one low-mass and one high-mass), initialized some constants, and defined the numerical integration loop structure.
We aimed to evaluate students' engagement of the modeling process by contextualizing a physics problem into a programming task. 
Furthermore, we intended to assess certain programming skills, namely, students' abilities to identify and assign variables and implement the numerical integration algorithm. 
The assignment was delivered using WebAssign in a timed mode (30 minutes) in each of the sixteen lab sections over eight meeting periods. 
TAs were not permitted to help students debug their programs, nor were students allowed any additional aids (Internet, book, notes, lab partners, et cetera).
Furthermore, the assignment was IP-restricted and password-protected; students could only access the assignment during their assigned lab time.
Even tough the assignment was proctored over 4 days, we believe it is unlikely that large number of students shared useful information about the assignment; scores for each lab section were statistically indistinguishable.
The format of the assignment was identical to students' final two homework problems; students were given a test case to check their solution before solving the grading case. 

For this assignment, students modeled the motion of the low-mass object as it interacted with the high-mass object through a central force. 
The nature of the force (attractive or repulsive) and its distance dependence ($r^n$) were randomized on a per-student basis. 
We also randomized some of the variable names in the partially-completed program  to hinder copying. 
After adding and modifying the necessary program statements, students ran their program and reported the final location and velocity of the low-mass object.
During the assignment, students did not receive feedback from the WebAssign system about the correctness of their answer, but they were given three attempts to enter it.
Similar to students' online homework, only the final numerical answer was graded.
These additional attempts afforded students who experienced browser issues or made typing mistakes the ability to resubmit their answers without penalty.

\begin{table}[t]
\caption{As part of a final proctored lab assignment, students completed a partially constructed program which modeled the motion of an object under the influence of a central force. The partially written program defined the objects, some constants and the numerical integration loop structure. The initial conditions, the sign ($\pm$) and distance dependence ($r^n$) of the force, and object names were randomized on a per-student basis. Slightly modified versions (Ver.) of this assignment were given at the end of three different semesters. Modifications were made to streamline delivery (Version 1 to Version 2), minimize transcription errors and improve presentation (Version 2 to Version 3). Students' performance on Version 1 was likely inflated because some students were allowed to work the problem on two separate occasions. \label{tab:eval}}
\begin{center}
\begin{tabular}{c|cc|c}
\multicolumn{1}{c}{\bf Ver.} & \multicolumn{1}{c}{\bf Correct} & \multicolumn{1}{c}{\bf Incorrect} & \multicolumn{1}{c}{\bf \% Correct}\\\hline\hline
1 & 303 & 168 & 64.3\\\hline
2 & 201 & 193 & 51.0\\\hline
3 & 316 & 176 & 64.2\\\hline\hline
\multicolumn{1}{c}{Overall} & \multicolumn{1}{c}{820} & \multicolumn{1}{c}{537} & \multicolumn{1}{c}{60.4}\\
\end{tabular}
\end{center}
\end{table}

Performance varied from semester to semester (Table \ref{tab:eval}) because the assignment was modified slightly between each semester in order to streamline delivery (Version 1 to Version 2), reduce transcription errors, and improve presentation (Version 2 to Version 3). 
In the first semester, students were permitted to attempt Version 1 of the assignment twice due to a logistical issue with the initial administration of the assignment. The majority of students (64.3\%) were able to model the grading case successfully on the second administration of the assignment. 
Students' performance on Version 1 was likely inflated because some students were able to work the problem twice. \footnote{The first administration of this assignment was during a regular hour exam. Roughly, 40\% of the students modeled the motion correctly. However, students used their own laptop computers which created several logistical challenges.} 
Students solved Version 2 only once, and student performance dropped.
A number of students were confused by the randomized exponent on the units of one of their initial conditions (Sec.  \ref{sec:errorfreq}).  
About half of the students (51.0\%) were able to model the grading case successfully. 
Students were more successful on Version 3 of the assignment; 64.2\% modeled the grading case correctly.

Overall, roughly 40\% of the students were unable to model the grading case. 
Students with some previous programming experience were no more successful than those without any experience as indicated by a contingency table analysis. \cite{numericalrecipes}
To determine exactly what challenges they faced while completing this assignment, we reviewed the program of each student who failed to model the grading case. 
Students who uploaded their programs to the WebAssign system received a small extra credit bonus on the proctored assignment.
Only a very small number of students ($<$1\%) did not upload any code.
We limited our review to the programs submitted for Versions 2 and 3 of the assignment.

\begin{table}[t]
\caption{Incorrectly written programs were subjected to an analysis using a set of codes developed from common student mistakes. The codes focused on three procedural areas: {\it using the correct given values} (IC), {\it implementing the force calculation} (FC) and {\it updating with the Newton's second law} (SL). We reviewed each of the incorrectly written student programs for each of the features listed below. These codes are explained in detail in Appendix \ref{sec:vpcodes}.}\label{tab:vpcodes}
\begin{center}
\begin{tabular}{ p{0.10\linewidth} p{0.85\linewidth} }\hline
\multicolumn{2}{p{0.95\linewidth}}{\bf Using the correct given values (IC)} \\
IC1 & Used all correct given values from grading case\\
IC2 & Used all correct given values from test case\\
IC3 & Used the correct integration time from either the grading case or test case\\
IC4 & Used mixed initial conditions\\
IC5 & Exponent confusion with $k$ (interaction constant)\\\hline

\multicolumn{2}{p{0.95\linewidth}}{\bf Implementing the force calculation (FC)} \\
FC1 & Force calculation was correct\\
FC2 & Force calculation was incorrect but the calculation procedure was evident\\
FC3 & Attempted to raise separation vector to a power\\
FC4 & Direction of the force was reversed\\
FC5 & Other force direction confusion\\\hline

\multicolumn{2}{p{0.95\linewidth}}{\bf Updating with Newton's second law (SL)} \\
SL1 & Newton's second law (N2) was correct\\
SL2 & Incorrect N2 but in an update form\\
SL3 & Incorrect N2 attempted update with scalar force\\
SL4 & Created new variable for $\vec{p}_f$\\\hline

\multicolumn{2}{p{0.95\linewidth}}{\bf Other (O)} \\
O1 & Attempted to update (force/momentum/position) for the massive particle\\
O2 & Did not attempt the problem\\\hline

\end{tabular}
\end{center}
\end{table}

\section{Systematically unfolding students' errors}\label{sec:vpcat}

Students must perform several tasks to successfully write and execute the program for the proctored assignment. 
Students must interpret the problem statement; that is, they must contextualize a word problem into a programming task. 
They must review the partially-completed program and identify the variables to update. 
Students need to apply their knowledge of predicting motion to the problem using VPython. 
They must identify that the force is non-constant, and then write the appropriate programming statements to calculate the vector force. 
Students need to then complete the motion prediction routine by writing a statement to update the momentum of the low-mass object. 

Using an iterative-design approach, we developed a set of binary (affirmative/negative) codes to check which tasks students performed correctly and which errors they made. 
An initial review of students' uploaded programs yielded the mistakes that were made most often. 
These common mistakes formed the basis for the codes. The codes were developed empirically, and several iterations were made before they were finalized. 
Two raters tested the codes by coding a single section of student submitted programs ($N = 45$). 
The raters resolved their differences (which further explicated the codes) and then recoded the section. 
The final codes (Table \ref{tab:vpcodes}) were used by both raters independently to code the remaining sections ($N = 324$). 
The final codes had high inter-rater reliability; both raters agreed on 91\% of the codes. 

We classified each code into one of three procedural areas: {\it using the correct given values} (IC), {\it implementing the force calculation} (FC) and {\it updating with the Newton's second law} (SL). 
These areas were congruent with the broad range of difficulties and misconceptions which students exhibited with their erroneous programs. 
Each code is explained in greater detail in Appendix \ref{sec:vpcodes}.

Determining where students encountered difficulties with these tasks might help explain how students learn this algorithmic approach to use Newton's second law to predict motion.  
Because we reviewed students' programs after they were written, we are unable to comment directly on students' challenges with contextualizing the problem. 
Our work was limited to analyzing students' procedural efforts (i.e., identifying variables and implementing the numerical integration algorithm). 
However, each of the authors has several years of experience with teaching computational modeling to introductory students from a wide variety of backgrounds. Hence, some information about students' thoughts and actions could be inferred from this analysis.

\section{Frequency of errors in students' programs\label{sec:errorfreq}}

We measured the frequency of students' errors within each category (IC, FC and SL) by mapping binary patterns extracted from our coding scheme to common student mistakes.
The number of possible binary patterns we observed in our data ranged from nine for SL to seventeen for FC with 13 possible for IC. 
Not all the codes within a given category were independent, hence, the number of possible binary patterns is much less than $2^n$.
Within a given category, we found that a large percentage of students could be characterized by just a few error patterns (between four and seven). 

The errors we observed were not necessarily unique to computational problems. 
The most notable errors involved calculating forces or updating the momentum. 
Most of these errors appeared to be physics errors reminiscent of those made on pencil and paper problems.
Many of them could have been mitigated by qualitative analysis. 
Some errors were unique to computational models and the iterative description of motion because they did not prevent the program from running, but nevertheless caused it to model the system inappropriately. 
Still others (e.g., replacing initial conditions) might have appeared to be simple careless mistakes; when investigated closely, however, these errors highlighted the fragility of students' knowledge.

\subsection{Initial Condition Errors}

\begin{table}[t]
\caption{Only seven of the fourteen distinct code patterns for the IC category (Table \ref{tab:vpcodes}) were populated by more than 3\% of the students. The patterns (ICx) are given by affirmatives (Y) and negatives (blank) in the code columns (IC\#). The percentage of students with each pattern is indicated by the last column (\%). These 7 patterns accounted for 88.8\% of students with erroneous programs.}
\begin{center}
\begin{tabular}{c|ccccc|c}
\multicolumn{7}{c}{\bf Initial Condition Codes} \\
\multicolumn{1}{c}{\bf Pattern} & {\bf IC1} & {\bf IC2} & {\bf IC3} & {\bf IC4} & \multicolumn{1}{c}{\bf IC5} & \multicolumn{1}{c}{\bf \%}\\\hline\hline
{\bf ICa} & & Y & Y & & & 27.6\\\hline
{\bf ICb} & & & Y & Y & & 16.0\\\hline
{\bf ICc} & Y & & Y & & & 14.4\\\hline
{\bf ICd} & & & Y & & Y & 13.8\\\hline
{\bf ICe} & & & & Y & & 7.9\\\hline
{\bf ICf} & & & Y & & & 5.2\\\hline
{\bf ICg} & Y & & & & & 3.8\\
\end{tabular}\label{tab:ic}
\end{center}
\end{table}

Students had to identify and update a total of eight given values: the interaction constant ($k$), the ``interaction strength'' ($n$), the mass of the less massive particle, the position and velocity of both particles, and the integration time.
Most students with incorrect programs (88.8\%) fell into one of seven IC patterns (Table \ref{tab:ic}). 
Students in ICa (27.6\%) did not use correct values from the grading case. 
Instead, these students identified and replaced all the initial conditions with those from the test case (IC2), including the integration time (IC3). 
Those in ICb (16.0\%) used a mixed set of initial conditions, that is, conditions from both the grading and test cases (IC4), and selected one of the two given integration times (IC3). 
Students in ICc (14.4\%) had a full set of correct initial conditions.
These students identified and correctly replaced all the initial conditions with those from the grading case (IC1), including the integration time (IC3). 
Students who appeared in ICd (13.8\%) used the correct integration time (IC3), but had a single initial condition mistake; they were confused by the exponent on the units of the interaction constant (IC5). 
Students in ICe (7.9\%) used a variety of initial conditions and given values (IC4). 
Those students in ICf (5.2\%) only identified and employed a given integration time (IC3); their initial conditions were completely incorrect.
Finally, students in ICg (3.8\%) used appropriate initial conditions from the grading case (IC1), but did not employ the grading case's integration time.
Most students might have simply forgotten to update one or more of the initial conditions from either the default case or the test case (ICb, ICe, ICf and ICg). 
A small fraction of students with mixed initial conditions had values from all three cases. 

Students in ICa were most likely stuck on the test case because they had trouble with another aspect of the problem. 
These students were unable to obtain the solutions provided in the test case, and so kept working on it.
It is possible a number of these students ran out of time while trying to debug their programs.

It is difficult to say definitively if students with mixed initial conditions (ICb and ICe) were unable to identify the appropriate values, as we reviewed students' programs only after they were submitted. 
It is possible that these students were just careless when making changes, but they might have been unable to identify and update these quantities. 
Some students could have been in the process of updating these quantities when they ran out of time and uploaded their programs.

Identifying and updating variables in a program is not a trivial task for students. 
In fact, students' difficulties with updating variables highlights the fragility of their computational knowledge.
As an example, consider the students who confused the exponent on the length unit of the interaction constant ($k$) for the exponent in the scientific notation of the numerical value of $k$ when they defined it in their programs (ICd). 
The distance dependence of the central force was randomized, and hence the units of the interaction constant ($k$) were dependent on a student's realization. 
In Version 2 of the assignment, the exponent on the length unit of $k$ was colored red (WebAssign's default behavior for random values). 
A student in ICd would read $k$= 0.1 Nm$^3$ to mean $k$ = 0.1E3 = 100 rather than $k$ = 0.1 {\it Newton times meters cubed}.
In Version 3 of the assignment, we changed the exponent's text color to black like the rest of the non-random text. 
The overall frequency of this mistake dropped from 30.5\% to 9.1\%.

\subsection{Force Calculation Errors}

Students were given the magnitude of the force as an equation ($F=kr^n$) and told that their (attractive or repulsive) force acted along the line that connected the two objects. 
In solving this problem, students had to correctly calculate the magnitude of the central force and identify the unit vector ($\hat{r}$) and sign ($\pm$) for their own realization. 
Almost all students (98.8\%) appeared with one of five FC patterns (Table \ref{tab:fc}). 
Students in FCa (23.9\%) implemented a reasonable force calculation procedure in the calculation loop (FC2), but reversed the direction of the net force (FC4). 
Those in FCb (22.2\%) performed the force calculation correctly (FC1). 
Students in FCc (15.7\%) implemented the procedure correctly (FC2), but were likely to include a force irrelevant to the problem (i.e., gravitational or electric interactions) or compute only the magnitude of the net force. 
Students who appeared in FCd (14.6\%) implemented the force calculation procedure correctly (FC2), but attempted to raise the separation vector to a power (FC3). 
Students in FCe (14.0\%) showed no evidence of an appropriate force calculation procedure (all codes negative); the procedure was either completely incorrect (e.g., used the differential form of the Impulse-momentum theorem) or was calculated outside the numerical integration loop (i.e., a constant force). 
Those students in FCf (8.4\%) had an appropriate force calculation procedure (FC2) but invented a unit vector for the net force (FC5).

\begin{table}[t]
\caption{Only six of the nine distinct code patterns for the FC category (Table \ref{tab:vpcodes}) were populated by more than 3\% of the students. The patterns (FCx) are given by affirmatives (Y) and negatives (blank) in the code columns (FC\#). The percentage of students with each pattern is indicated by the last column (\%). These 6 patterns accounted for 98.8\% of students with erroneous programs.}
\begin{center}
\begin{tabular}{c|ccccc|c}
\multicolumn{7}{c}{\bf Force Calculation Codes} \\
\multicolumn{1}{c}{\bf Pattern} & {\bf FC1} & {\bf FC2} & {\bf FC3} & {\bf FC4} & \multicolumn{1}{c}{\bf FC5} & \multicolumn{1}{c}{\bf \%}\\\hline\hline
{\bf FCa} & & Y & & Y & & 23.9\\\hline
{\bf FCb} & Y & & & & & 22.2\\\hline
{\bf FCc} & & Y & & & & 15.7\\\hline
{\bf FCd} & & Y & Y & & & 14.6\\\hline
{\bf FCe} & & & & & & 14.0\\\hline
{\bf FCf} & & Y & & & Y & 8.4\\
\end{tabular}\label{tab:fc}
\end{center}
\end{table}

The difficulties that students faced when numerically computing the net force likely stems from a weak grasp of the concept of vectors. 
Students in FCa made directional mistakes (e.g., changing the sign of one of lines 21--23 in Fig. \ref{fig:hwsamplecode}) that could have been easily identified and rectified by drawing a sketch of the situation, a problem-solving strategy that is practiced in the laboratory. 
Those who raised the separation vector to a power (FCd) likely transcribed the central force equation (replacing $r$ by $\vec{r}$) without thinking that this operation was mathematically impossible ($\vec{F} \sim k(\vec{r})^n$ vs. $\vec{F} \sim k|\vec{r}|^n\hat{r}$). 
We have found that students attempt a similar operation on pencil and paper problems; raising components of a vector to a power (e.g., $(\vec{r})^n = \langle r_x^n, r_y^n, r_z^n \rangle$). 
Such mistakes are quite common among introductory physics students. \cite{knight1995vector}
However, in the pencil-and-paper case, students are not immediately directed to their mistake as they are in a programming environment.
VPython raises an exception error when this operation is attempted. 
These students appeared to be unable to parse this error into any useful information. 
Students who make this type of error might be helped by additional exposure to translating force equations to precise programmatic representations. \cite{sherin1993dynaturtle}
Some students invented a unit vector (FCf) for the net force. 
This was most likely because they had computed a scalar force and tried to add a scalar impulse to the vector momentum. 
VPython raises a different exception error upon any attempt to add a scalar to a vector. 
These students were able to parse this error, but resolved it incorrectly.

Other students (FCc) might have incorrectly contextualized the problem by including an irrelevant force (i.e., gravitational or electrical interactions). 
The problem clearly stated that the two objects were far from all other objects. 
It did not explicitly state to neglect the gravitational interaction between the objects. 
However, the gravitational interaction could be safely neglected for the range of masses and distances we had chosen. 
Furthermore, nothing about the charge of the objects was mentioned in the problem statement. 
It is surprising that students included these interactions in their models. 
One possible explanation for the inclusion of these interactions is that students had memorized how to solve the gravitational and Coulomb problems because these problems had appeared on their homework several times and on an exam. 
They might have panicked and simply wrote all possible forces they could remember.
We have observed similar contextualization issues in laboratory instruction and on hand-written exams.

A number of students (FCe) did not employ the force calculation algorithm at all. 
Some of these students computed the net force (e.g., lines 21--23 in Fig. \ref{fig:hwsamplecode}) outside the numerical integration loop (e.g., before line 19 in Fig. \ref{fig:hwsamplecode}). 
In this case, the net force was effectively constant and therefore only correct at $t=0$ . 
A program with correct syntax will run regardless of its physical implications.
This error is unique to computational problems in which motion is predicted iteratively. 
Students in introductory physics rarely use Newton's second law to predict the motion of objects that experience non-constant forces.
Other students who fell into FCe wrote ``creative'' program statements. 
Students in this group manipulated some quantities in the loop but did not perform any physically relevant calculations. 
The number of students with ``creative'' program statements was relatively small.

\subsection{Newton's Second Law Errors}

Students had to write a program statement similar to line 25 in Fig. \ref{fig:hwsamplecode} to properly update the momentum using Newton's second law. 
Most students demonstrated no difficulty in remembering the formula for the momentum update but some met challenges with making that description precise. \cite{sherin1993dynaturtle} 
Nearly all students (95.7\%) fell into one of four SL patterns (Table \ref{tab:mp}). 
Most students appeared in SLa (69.7\%) because they wrote Newton's second law in the correct iterative format, {\tt p = p + F*dt} (SL1).
A much smaller number of students fell into SLb (13.2\%).
These students wrote Newton's second law in an iterative format (SL2), but they attempted to update the vector momentum with a scalar force (SL3). 
Students in SLc (7.9\%) were unable to write Newton's second law in any form that updated (all codes negative). 
A small fraction (SLd, 4.9\%) wrote Newton's second law in an iterative form (SL2), but did so incorrectly (e.g., by creating a new variable). 

\begin{table}[t]
\caption{Only four of the nine distinct code patterns for the SL category (Table \ref{tab:vpcodes}) were populated by more than 3\% of the students. The patterns (SLx) are given by affirmatives (Y) and negatives (blank) in the code columns (SL\#). The percentage of students with each pattern is indicated by the last column (\%). These 4 patterns accounted for 95.7\% of students with erroneous programs.}
\begin{center}
\begin{tabular}{c|cccc|c}
\multicolumn{6}{c}{\bf Second Law Codes} \\
\multicolumn{1}{c}{\bf Pattern} & {\bf SL1} & {\bf SL2} & {\bf SL3} & \multicolumn{1}{c}{\bf SL4} & \multicolumn{1}{c}{\bf \%}\\\hline\hline
{\bf SLa} & Y & & & & 69.7\\\hline
{\bf SLb} & & Y & Y & & 13.2\\\hline
{\bf SLc} & & & & & 7.9\\\hline
{\bf SLd} & & Y & & & 4.9\\
\end{tabular}\label{tab:mp}
\end{center}
\end{table}

Students who attempted to update the momentum with a scalar force (SLb) might have been facing difficulties with understanding vectors. 
The momentum update was presented as a vector equation ($\vec{p}_f = \vec{p}_i + \vec{F} \Delta t$). 
These students might have been unable to unpack that representation into a precise programmatic description, but it was more likely that they calculated a scalar force (FCc) and then simply wrote the correct (vector) second law syntax.
VPython raises an exception error if an attempt to add a vector to a scalar is made. 
The students appeared unable to parse this error into any useful information.

Students who were unable to write Newton's second law in any form that updated  (SLc) might have experienced difficulties with converting the second law formula into a precise and useful programmatic representation.
Students in this category either wrote Newton's second law in a non-update form (e.g. writing {\tt deltap = Fnet*deltat} or {\tt pf - pi = Fnet*deltat} as line 25 in Fig. \ref{fig:hwsamplecode}), or wrote a number of program statements which manipulated quantities but performed no useful calculations. 
In either case, these students could benefit from the precision required by a programming language.\cite{sherin1993dynaturtle} By requiring them to accurately represent Newton's second law in their programs, they might begin to distinguish between the utility and applicability of its various algebraic forms.

Students who wrote Newton's second law in a form that updated incorrectly (SLd) either remembered the formula for the second law incorrectly or made a typo.
These students would generally leave off the timestep in the momentum update (e.g., {\tt p = p + F}) or divide by it ({\tt p = p + F/deltat}). 
Dividing by the timestep is a particularly egregious error because the numerical value of the timestep was quite small. Hence, the impulse added in this case would have been huge. 
Students who made this error were unable to assess the state of the visualization (the particle flew off to ``infinity'') to debug this error.

\section{Common Error Patterns in Students' Programs}\label{sec:cluster}

The patterns within individual categories (IC, FC, and SL) indicated the frequency of common mistakes students made when solving the proctored assignment, but a single student could make one or more of these mistakes. 
Evaluating a student's complete solution requires an analysis using all the codes (Table \ref{tab:vpcodes}).
In principle, the codes we developed could have had up to $\sim4300$ possible error patterns using all sixteen codes.
In fact, the intersections of code categories indicated that the number of distinct errors made by students across all categories was relatively small; we found only 111 distinct binary patterns.
It is possible to relate these unique patterns in a manner that suggests dominant  common errors.

Cluster analysis, a technique borrowed from data mining, is particularly well suited for this application because it characterizes patterns in complex data sets.\cite{tan2006introduction, clustereveritt} 
This technique has been used previously to classify students' responses to questions about acceleration and velocity in two dimensions.\cite{springuel2007applying} 
It was used here to determine the major features in students' incorrect programs which were responsible for their failure. 

We applied the cluster analysis technique to the data generated from our set of binary codes. 
We used the Jaccard metric \cite{jaccard1901study} to measure inter-cluster distances and linked clusters using their average separation.\cite{sokal1975statistical} 
We tested several other metrics (e.g., Hamming, city block, etc.). 
The Jaccard metric was chosen because it neglects negative code pairs. 
Both the Hamming and city block metrics produced similar pairings at low levels, but higher order clusters were difficult to interpret.
We used average linkage to avoid the effects of ``chaining'' that appeared when nearest neighbor linkage \cite{sneath1957application} was used and because useful clusters were more difficult to distinguish when farthest neighbor linkage \cite{sorenson1948method} was used.

Thirty clusters with inter-cluster distances below 0.5 were reviewed in detail. This cutoff was selected to minimize the number of unique clusters while still rendering clusters with useful interpretations. 
Most students (86.5\%) appeared in seven of the thirty clusters (Table \ref{tab:clus}). 
These clusters had very few students ($<$1\%) with affirmatives in the ``Other'' category. 
Codes O1 and O2 were dropped from Table \ref{tab:clus} for this reason. 
Each of the other 23 clusters were populated by less than 3\% ($N \approx 10$) of the students, and the bottom 18 clusters had less than 1\% ($N \approx 3$) each.
Each of the dominant clusters demonstrated a unique challenge that students faced while solving the proctored assignment (Table \ref{tab:clus}). 

\begin{table*}[t]
\caption{Only seven of the thirty clusters with an inter-cluster distance of less than 0.5 were populated by more than 3\% of the students. The bottom 18 clusters were populated by less than 1\% of students each. These seven clusters accounted for 86.5\% of students. The percentage of affirmatives for each code (Table \ref{tab:vpcodes}) within any given cluster (A-G) is given to the nearest whole percentage. Codes with affirmative percentages greater than 60\% are bolded. These clusters had very few students ($<1$\%) with any affirmatives in the `Other'' category, hence the results from this category are not reported. The percentage of students in each cluster is indicated in the last column (\%). 
}
\begin{center}
\begin{tabular}{c|ccccc|ccccc|cccc|c}
\multicolumn{1}{l}{\hspace*{1pt}} & \multicolumn{5}{c}{\bf Initial Conditions} & \multicolumn{5}{c}{\bf Force Calculation} & \multicolumn{4}{c}{\bf Second Law} & \multicolumn{1}{l}{\hspace*{1pt}}\\
\multicolumn{1}{c}{\bf Cluster} & {\bf IC1} & {\bf IC2} & {\bf IC3} & {\bf IC4} & \multicolumn{1}{c}{\bf IC5} & {\bf FC1} & {\bf FC2} & {\bf FC3} & {\bf FC4} & \multicolumn{1}{c}{\bf FC5} & {\bf SL1} & {\bf SL2} & {\bf SL3} & \multicolumn{1}{c}{\bf SL4} & \multicolumn{1}{c}{\bf \%}\\\hline\hline
A & 00 & {\bf 68} & {\bf 93} & 18 & 15 & 00 & {\bf 100} & 22 & {\bf 66} & 09 & {\bf 95} & 00 & 00 & 01 & 23.8\\\hline
B & 21 & 01 & {\bf 86} & 37 & 41 & {\bf 88} & 00 & 00 & 00 & 00 & {\bf 97} & 00 & 00 & 00 & 19.8\\\hline
C & 04 & 33 & {\bf 76} & 31 & 22 & 00 & {\bf 94} & 00 & 08 & 00 & 00 & {\bf 98} & {\bf 98} & 08 & 13.3\\\hline
D & {\bf 98} & 00 & {\bf 85} & 00 & 00 & 00 & {\bf 85} & 18 & 50 & 00 & {\bf 98} & 00 & 00 & 00 & 10.8\\\hline
E & 00 & 00 & 57 & {\bf 75} & 36 & 00 & {\bf 100} & {\bf 79} & 00 & 04 & {\bf 89} & 00 & 00 & 00 & 7.6\\\hline
F & 00 & {\bf 100} & {\bf 96} & 00 & 00 & {\bf 65} & 00 & 00 & 00 & 00 & {\bf 73} & 19 & 00 & 04 & 7.1\\\hline
G & 27 & 00 & {\bf 93} & 53 & 07 & 00 & {\bf 100} & 00 & 07 & {\bf 100} & {\bf 93} & 00 & 00 & 00 & 4.1\\
\end{tabular}\label{tab:clus}
\end{center}
\end{table*}

Students in cluster A (23.8\%) tended to remain stuck on the test case (ICa) due to an error in their force calculation. Reversing the direction of the force (FCa) was the most common mistake, followed by raising the separation vector to a power (FCd). Most students in this cluster had no trouble expressing Newton's second law (SLa).
These students worked diligently to solve the test case but were unable to do so. As a result, they did not proceed to the grading case.

Cluster B (19.8\%) contained students who made mistakes while replacing the given values and initial conditions (any IC code except ICa). 
Some of these students worked with the grading case (ICc and ICg).
Others might have been working with either case and had mixed conditions (ICb and ICe) or simply incorrect ones (ICf).
Still others might have incorrectly assigned the exponent on the length unit of k to the exponent of k in scientific notation (ICd).
At any rate, most students in this cluster were able to construct a working albeit incorrect program. 
Given their unfamiliarity with general central force interactions, these students might have believed their solutions were correct. 
In fact, it is possible that students who were working with the grading case (ICc and ICg) had solved the test case correctly and simply made a typo.

Students in cluster C worked with either the grading or test case and might have made a number of mistakes with their initial conditions (any IC code except ICa).
The dominant error in cluster C was committed by students who computed the magnitude of the net force (FCc) and attempted to update the vector momentum with this scalar force (SLb). 
This mathematically impossible operation raises a VPython error. Students in this cluster were unable to parse this error into any useful information.

Cluster D (10.8\%), like cluster A, was populated by students who tended to make errors in the force calculation (FCa and FCd), but students in Cluster D worked with the grading case (ICc). 
The most common error in Cluster D was reversing the direction of the net force (FCa), followed by raising the separation vector to a power (FCd). 
Again, like cluster A, most students met no challenges when updating the momentum using Newton's second law (SLb). 
These students might have started working with the test case, but we think it is more likely that they started working with the grading case because the dominant error appears in their force calculations.
We have observed a number of students doing so in homework help sessions.

Students in cluster E (7.6\%) tended to raise the separation vector to a power (FCd) and have mixed initial conditions (ICb, ICd, ICe and ICf). 
These students generally had no difficulty with writing Newton's second law correctly (SLa). The dominant error for students in cluster E was raising the separation vector to a power (FCd). This mathematically impossible operation raises a VPython error. Students in this cluster were unable to parse this error into any useful information.

Cluster F (7.1\%) contained students who worked solely with the test case (ICa) and either had no issue with their force calculation (FCb) or had no evident force calculation procedure (FCe). 
Most of these students had no difficulty updating the momentum using Newton's second law (SLa).
Students in cluster F were able to construct a program which ran without raising any VPython errors. 
Students who had no issue with their solution likely completed test case but simply ran out of time before turning to the grading case. 
Students with no evident procedure generally computed the net force outside the numerical calculation loop, essentially making this force constant in time. 
Given students' unfamiliarity with general central force interactions, it would not be surprising if students who treated the central force outside the loop believed their solutions were correct.

Students in cluster G (4.1\%) all invented an incorrect unit vector for the force rather than using $\hat{r}$ (FCf) regardless of the case with which they worked (ICc,  ICb, and ICf).
These students generally had no difficulty updating the momentum using Newton's second law (SLa). 
Most likely, these students computed the magnitude of the force (similar to students in cluster C) but were able to parse the resulting VPython error. 
Students in cluster G corrected their mistake by assigning some unit vector to the force before the momentum was updated. 

\section{Closing remarks}\label{sec:closing}

Students with little to no programming experience can, in large introductory physics courses, develop the computational skills necessary to predict the motion of sundry physical systems. 
After solving a suite of computational homework problems, most students ($\sim 60$\%) were able to model the motion of a novel problem successfully.
In our work, we discovered that most students who were unsuccessful encountered challenges when calculating the net force acting on the object in the motion prediction algorithm (Clusters A and C through G in Table \ref{tab:clus}).
By contrast, there were fewer students whose primary challenge was identifying and assigning variables (Cluster B in Table \ref{tab:clus}). 
We acknowledge that we have limited the development of our students' computational skill set to contextualizing a word problem into a programming task, identifying and updating input variables, and applying a motion prediction algorithm. 
We believe that further development of our homework problems and other novel deployments could broaden the scope of the skills students develop.

Our work provides a touchstone example of computational instruction for the STEM education community. This work provides clear evidence that newcomers to both computer programming and computational modeling can successfully acquire, in a large lecture introductory STEM course, basic skills needed to do computational modeling. However, the success of our students on the proctored assignment should not be overemphasized. Ours is but one of many possible implementations that might be successful in a large lecture setting. Moreover, the level of instruction in this course was limited to the iterative prediction of motion. What is more fascinating are the types of errors which students made when solving the proctored assignment and the commonalities among those errors.

Procedural errors such as those we have documented (Secs. \ref{sec:errorfreq} \& \ref{sec:cluster}) could be corrected by developing additional materials aimed at addressing each error individually. 
However, the results from this work indicate that instructional efforts should be focused not only on correcting procedural mistakes but also on developing students' qualitative habits of mind. 
Training students to write programs to predict motion might help them to be successful in a highly structured environment, but they would be better served by learning the practice of debugging. 
Here, debugging includes identifying syntax errors (of which we found few) and, more importantly, performing the type of qualitative analysis that is typically taught for solving analytic problems. 
Students who could synthesize their analytic and computational skills would be better prepared to solve the open-ended problems they will face in their future work. 
Students' use of debugging is the subject of active investigation by the computer science education community.\cite{McCauley2008,Fitzgerald2008,Murphy2008}

Developing the materials to teach these skills requires an evaluation of how students contextualize computational problems. 
We do not claim to understand this contextualization presently, although we have been able to glean some suggestive information based on students' errors. 
Investigating what students think about when solving computational problems requires structured student interviews (i.e., a think-aloud study). 
In the future, we plan to perform such a study to not only characterize students' abilities to contextualize but also to elucidate the mechanism for some of the errors we reported in Secs. \ref{sec:errorfreq} \& \ref{sec:cluster}.
Moreover, this work provides the foundation for others in the STEM community to perform similar investigations.

Research into skill development in math and science has shown a strong correlations with student epistemology.\cite{schommer1990effects,schommer1993epistemological} 
Epistemology is important because the views that students hold affect how they learn \cite{elby2001helping} and, ultimately, how successful they are in their science courses.\cite{halloun1997views,perkins2005correlating} 
It is therefore crucial that we understand students' sentiments about learning a new tool such as computation. 
Our students expressed anxiety and demonstrated a lack of self confidence in laboratory instruction and homework help sessions, even with their additional exposure to computational problems and improved scores on hand-written programming questions. 
We are developing an attitudinal survey aimed at exploring these and other beliefs in detail.
Students who learn to use computational modeling and are confident in their abilities will be better prepared to solve challenging problems.

We do not claim to have assessed a transfer of computational knowledge. 
We designed a set of problems (Sec. \ref{sec:vpdeploy}) that students solved over the course of the semester with an eye toward a final assessment of their skills using a novel problem. 
This problem (Sec. \ref{sec:eval}) was similar to some of the homework problems students had solved previously. 
It focused on key skills that we desired students to acquire: contextualizing a problem, identifying and assigning variables in a program, and carrying out the motion prediction algorithm.
An evaluation of transfer would require that students apply these computational skills to a different domain (e.g., electromagnetism) or a different task (e.g., open-ended inquiry). 
Demonstrating transfer of computational knowledge is a necessary step in developing students into flexible problem solvers for the 21st century.

It is the goal of many reforms in physics education to develop students into flexible problem solvers while exploring the practice of science. 
Teaching computational modeling alongside physics provides support for that effort. 
Students learn the tools for doing science while developing a qualitative understanding of physical systems, exploring the generality of physics principles, and learning broadly applicable problem solving methods in computation. 

\begin{acknowledgments}
The authors appreciate the helpful comments from the two anonymous reviewers who helped focus and improve the paper. 
We would also like to thank Edwin Greco for help in deploying our computational problems, Balachandra Suri for analyzing some of the student-written code, and Scott Douglas for carefully reviewing and editing the manuscript.
This work was supported by National Science Foundation's Division of Undergraduate Education (DUE0618519 and DUE0942076).
\end{acknowledgments}

\appendix
\section{More details on the evaluation codes} \label{sec:vpcodes}

The codes shown in Table \ref{tab:vpcodes} were developed empirically. The procedure followed an iterative-design approach. We reviewed student work for common errors and devised a rough coding scheme. We then tested the scheme on a new set of student submitted programs. The scheme was refined and re-tested. This iterative procedure was repeated several times until we captured the majority of students' mistakes. Each code is explained in detail below.

\subsection{Using the correct given values (IC) Codes}

We reviewed the variables in each student's program. The default values had to be updated with the values given in the problem statement in the partially-completed program. We present the codes used to categorize each student's program with respect to identifying and updating the appropriate initial conditions for their realization.

\paragraph*{IC1 -- Student used all the correct given values from grading case.}
A student must replace the values of all the variables (mass, position, and velocity, interaction constant $k$ and the exponent in the force law $n$ in $F=kr^n$) with those given in the {\it grading case}. This code excluded the integration time. It was intended that the larger mass object was to remain at its location. This was made explicit in the problem statement; the initial position $\langle 5,4,0 \rangle$ m and velocity $\langle 0,0,0 \rangle$ m/s of the larger mass of object were given in the problem statement, even though these same values appeared in the partially-written program.

\paragraph*{IC2 -- Student used all the correct given values from test case.}
A student must replace the values of all the variables (mass, position, and velocity, interaction constant $k$ and the exponent in the force law $n$ in $F=kr^n$) with those given in the {\it test case} for this code to apply. This code excluded the integration time. It was intended that the larger mass object was to remain at its location (See IC1). 

\paragraph*{IC3 -- Student used the correct integration time from either the grading case or test case.}
A student must replace the default integration time (1 s) with the values given in the case with which they intended to work (grading or test). A student who mixed initial conditions was given an affirmative on this code if the majority of their initial conditions were from the same case as the integration time.

\paragraph*{IC4 -- Student used mixed initial conditions.}
A student who used some but not all of the initial conditions from any of the cases (default, test, or grading) was given an affirmative on this code. This code excluded the integration time.

\paragraph*{IC5 -- Students confused the exponents on the units the exponent of $k$ (interaction constant).}
Many students incorrectly thought the exponent on the length unit of the interaction constant was the scientific notation exponent for the interaction constant itself. For example, a student thought $k = 0.1$ Nm$^3$ meant $k = 100$ rather than $k = 0.1$ {\it Newton times meters cubed}.

\subsection{Implementing the force calculation (FC) Codes}

We reviewed how the students employed the force calculation algorithm in each of the programs written for the proctored assignment. The partially-written program given to the students left out all statements related to the force calculation. Students were required to fill in this procedure using the appropriate VPython syntax. We present the codes used to categorize each student's program with respect to computing the vector force acting on the low-mass object.

\paragraph*{FC1 --The force calculation was correct.}
For this code to apply, a student must compute the separation vector, its magnitude, its unit vector, the magnitude of the force, and the vector force correctly. Each of these steps may be combined as long as the final result computes the vector force acting on the less massive particle at each instant. These steps must all appear in the numerical integration loop.

\paragraph*{FC2 -- The force calculation was incorrect, but the calculation procedure was evident.}
In the numerical integration loop, the student must perform a position vector subtraction, a calculation of the force magnitude and some attempt at combining magnitude with unit vector (any unit vector was acceptable) for this code to apply. If a student treated the problem using components and had some force which is a vector, it was coded as evident. If any part of the calculation was performed outside the loop, it was coded as {\it not} evident.

\paragraph*{FC3 -- The student attempted to raise the separation vector ($\vec{r}$) to a power.}
Students who raised the separation vector to a power generated a VPython exception error:\\
{\tt unsupported operand type(s) for ** or pow(): 'vector' and 'int'}.\\
This error told them that VPython cannot raise a vector to a power, as it is a mathematically impossible operation.

\paragraph*{FC4 -- The direction of the force was reversed.}
Students had to assign the correct unit vector and sign to the force depending on whether their force was attractive or repulsive for this code to be used. This code was not used if the student calculated the force as a magnitude only, raised $\vec{r}$ to a power, or invented a unit vector (e.g., $\langle 1,0,0 \rangle$). Visual feedback (i.e., the lower mass particle flying off to infinity) indicated a simple sign mistake.

\paragraph*{FC5 -- Student had some other force direction confusion.}
Some students used vectors other than $\vec{r}$ or $-\vec{r}$ to compute $\vec{F}$. Other students computed the force as a magnitude and then multiplied it by an ``invented'' unit vector (e.g., $\langle 1,0,0 \rangle$, $\hat{p}$). Both of these errors were given an affirmative for this code.

\subsection{Updating with the Newton's second law (SL) Codes}

We reviewed how the students employed the momentum update in each of the programs written for the proctored assignment. The partially written program given to the students left out the one line of code necessary to update the momentum. Students were required to fill in this line using the appropriate VPython syntax. We present the codes used to categorize each student's program with respect to updating the momentum of the low-mass particle.

\paragraph*{SL1 -- Newton's second law was correct.}
Correct Newton's second law meant that it was ``correct as a physics principle'' and also that it appeared ``in the update form''. This meant that {\tt pfinal = pinitial + Fnet*deltat} alone in a loop did not fall under ``correct Newton's second law''. It is an incorrect update form.

\paragraph*{SL2 -- Newton's second law was incorrect but in a form that updates.}
Newton's second law updates the momentum, but not necessarily correctly. (e.g., \texttt{p = p + Fnet}, \texttt{p = p + Fnet/dt}, \texttt{pf = p + Fnet}, etc. )

\paragraph*{SL3 -- Newton's second law was incorrect and the student attempted to update it with a scalar force.}
Some students computed the magnitude of the force acting on the particle and then used this magnitude to update the momentum. Students who did this raised a VPython exception error:\\
{\tt unsupported operand type(s) for +: 'vector' and 'int'}\\
This might have lead some to invent unit vectors in the momentum update (e.g., {\tt p = p + vector(Fmag,0,0)*dt}).

\paragraph*{SL4 -- Student created a new variable for $\vec{p}_f$.}
In computational modeling, the equal sign in a update line means ``add and replace''. Some students used a new symbol for the final momentum (e.g. {\tt pfinal}) and then replaced the momentum in the next step (e.g. {\tt p = pfinal}). Others only did the former, that is, they did not replace the momentum with its updated value.

\subsection {Other Codes}

Two common errors were not included in the above codes because they do not reflect errors in the procedure of modeling the motion of the low-mass particle. We present two miscellaneous codes which were common enough to consider relevant.

\paragraph*{O1 -- Student attempted to update force, momentum or position for the massive particle.}
The massive particle was intended to remain in place.

\paragraph*{O2 -- Student did not attempt the problem.}
Some students uploaded blank text files. We assumed they did not attempt solving the problem, and intended only to receive bonus credit for uploading their code.

\bibliography{vp}
\bibliographystyle{apsper}

\end{document}